\documentclass[aps,prl,twocolumn, showpacs]{revtex4}
\usepackage{graphicx}
\usepackage{dcolumn}
\usepackage{amsmath}

\begin{document}

\title{Electron-Dephasing Time in A Two-Dimensional Spin-Polarized System with Rashba Spin-Orbit Interaction}
\author{D. C. Marinescu}
\affiliation{Department of Physics and Astronomy, Clemson
University, 29634, Clemson }

\date{\today}
\begin{abstract}
We calculate the dephasing time $\tau_{\phi}(B)$ of an electron in a
two-dimensional system with a Rashba spin-orbit interaction,
spin-polarized by an arbitrarily large magnetic field parallel to
the layer. $\tau_\phi (B)$ is estimated from the logarithmic
corrections to the conductivity within a perturbative approach that
assumes weak, isotropic disorder scattering. For any value of the
magnetic field, the dephasing rate changes with respect to its
unpolarized-state value by a universal function whose parameter is
$2E_Z/E_{SOI}$ ($E_Z$ is the Zeeman energy, while $E_{SOI}$ is the
spin-orbit interaction), confirming the experimental report
published in Phys. Rev. Lett. {\bf 94}, 186805 (2005) . In the high
field limit, when $2 E_Z >> E_{SOI}$, the dephasing rate saturates
and reaches asymptotically to a value equal to half the
spin-relaxation rate.
\end{abstract}
\maketitle
\section{Introduction}
In two-dimensional electron gases (2DEG) with spin-orbit interaction
(SOI), the skew-scattering of the electron spin leaves a trademark
fingerprint on the quantum corrections to the conductivity in the
form of a positive contribution, the antilocalization term. This
results from the SOI mediated coupling of trajectories of electrons
with opposite momenta and opposite spins, a configuration aptly
named a singlet Cooperon, in the impurity-averaged diffusion
equation \cite{pikus}. The phase difference in the interfering paths
of the two electrons is measured by the dephasing time $\tau_\phi$.

A magnetic field of intensity $B$ applied parallel to the 2D layer
diminishes the spin-scattering effect
 of SOI by aligning the spins parallel to its direction.
 In this situation, $\tau_\phi (B)$ is
expected to depend on two relevant parameters: the Zeeman splitting,
proportional with the effective gyromagnetic factor $\gamma$,
$E_Z=2\gamma B$  and the spin-orbit energy, expressed in terms of
the spin-relaxation time $\tau_s(0)$ \cite{dyakonov}, $E_{SOI}
  = \hbar/\tau_s(0)$.

A theory of this effect, developed at low magnetic fields by
considering the Zeeman interaction as a perturbation on the electron
diffusion equation \cite{pikus}, finds that, for a SOI with linear
(Rashba \cite{rashba}) momentum coupling, $\tau_\phi(B)$ is
proportional to $(E_Z/E_{SOI})^2$ \cite{malshukov}. This behavior
has been confirmed by several separate experiments
\cite{meijer1,minkov}.

 The linear $(E_Z/E_{SOI})^2$ dependence of $\tau_\phi(B)$ inferred from
the low field estimates has been challenged recently by data
obtained by Meijer et al. \cite{meijer} for intense magnetic fields.
When $E_Z/E_{SOI}>>1$ the dephasing time saturates, becomes
independent of such system parameters as the electron density,
Rashba splitting, or the elastic scattering time, and is represented
by a universal function of $(E_Z/E_{SOI})^2$.

Challenged by these experimental findings, we calculate the
dephasing time and study its variation with $E_Z$ from a general
theory of the localization effects in a spin-polarized 2DEG with
Rashba spin-orbit coupling. The first important result of this paper
is that within the weak, isotropic scattering approximation,
$\tau_\phi(B)$ is, indeed, a universal function of
$(2E_Z/E_{SOI})^2$.  This dependence, rather than the
$(E_Z/E_{SOI})^2$ parametrization discussed in the experiment, might
be explained by the argument that the additional dephasing
introduced by the longitudinal magnetic field in the trajectories of
the opposite spin electrons that form the singlet Cooperon is such
that it preserves the relative orientation of the two spins, making
$2E_Z$ (corresponding to two spin flips) the appropriate energy
scale. The second important result is that in the high-field limit,
 the calculated
relative change of the dephasing rate in respect to its unpolarized
value tends asymptotically to $1/2\tau_s(0)$. Of course, at low
fields we recover the dephasing rate obtained in
Ref.~\onlinecite{malshukov}.

 The object of this study is a 2DEG, spin-polarized by a magnetic field parallel to the
 layer, $\vec{B} = B\hat{z}$. The magnetic field determines an imbalance between the
 number of spins parallel to the field, $n_\uparrow$ and those
 opposite to the field, $n_\downarrow$.
 The degree of spin polarization, $\zeta =
 (n_\uparrow-n_\downarrow)/(n_\uparrow+n_\downarrow)$,
 varies continuously from $-1$ to $1$ as a
 function of $B$ and is considered a parameter of the problem. The direction of the magnetic
 field, in the plane of the gas, corresponds to the quantization axis of the electron spin, $\hat{z}$ ($\hat{x}$ is in
 the plane, while $\hat{y}$ is perpendicular on the plane). The
 non-interacting electron energies, in the absence of the spin-orbit
 interaction, reflect the Zeeman splitting and are written, for an electron of momentum $\vec{p}$,
 spin $\sigma$, and effective mass $m^*$ (considered to be spin independent),
  as $\epsilon_{\vec{p},\sigma} = p^2/2m^* -\gamma \sigma B$, where $\sigma =1$ for
 a spin parallel to the field and $\sigma = -1$ for the opposite.

  In the presence of the Rashba
 spin-orbit coupling, the single-electron Hamiltonian is
\begin{equation}
H_{SOI} = \frac{\vec{p}^2}{2m} -
\frac{\tilde{\lambda}}{\hbar}\vec{\sigma}\cdot(\vec{p}\times
\hat{y})-\gamma \vec{\sigma}\cdot \vec{B}\;,
\end{equation}
with $\tilde{\lambda}$ the coupling constant of the Rashba
interaction, and $\vec{\sigma}$ the Pauli spin operator.

To calculate the dephasing time, we generalize the well known theory
of localization \cite{altshuler} for a spin-polarized system. By
considering the spin polarization as a preexisting condition of the
problem, rather than a perturbation in comparison with SOI, as in
Ref.~\cite{malshukov}, we allow an unconstrained relationship
between $E_Z$ and $E_{SOI}$ to occur. The
 logarithmic corrections to the conductivity are generated by the poles of a quantum diffusion equation, or, in
 an
 equivalent representation \cite{larkin}, by the eigenvalues
 of the Cooperon equation.
The propagator (or Cooperon) represents the convolution in the
momentum space of two single-particle Green functions, averaged over
the impurity configuration. To construct the propagator, we start by
first obtaining an expression for the single-electron Green function
in a spin-polarized electron system with Rashba interaction.

In the absence of SOI, the single particle Green function in the
2DEG is represented by a diagonal $2\times 2$ matrix in spin space:
 \begin{equation}
G^{R,A}(\vec{p},\omega) = \left(\begin{array}{cc}
                       G^{R,A}_\uparrow(\vec{p},\omega)&0\\
                       0&G^{R,A}_\downarrow(\vec{p},\omega)
                       \end{array}
                       \right)
                       \;,
                       \end{equation}
                       where
$
 G^{R,A}_\sigma(\vec{p},\omega)=(\omega - \hbar^{-1}\epsilon_{\vec{p}\sigma}
 \pm
 \frac{i}{2\tau_0})^{-1}
$ are the spin-dependent retarded (R) (with +) and advanced (A)
(with $-$) components. The impurity scattering is assumed to be
 isotropic, characterized by a rate $\tau_0 = 2\pi N_0 u^2$,
 with $N_0 = m/2\pi \hbar^2$ the single spin density of states at
 the Fermi surface, and $u^2$ the mean-square impurity potential, considered spin-independent.

The non-averaged propagator, ${ P}^0(\vec{q}, \Omega) $ defined by
\begin{equation}
{ P}^0(\vec{q}, \Omega) = \int d\omega \sum_{\vec{p}}
{G}^R(\vec{p},\omega){ G}^A(-\vec{p}+\vec{q},\omega+\Omega)\;,
\label{eq:prop0}
\end{equation}
involves two Green functions associated with two different spin
$1/2$ particles. Its matrix representation occurs in the $S_1\otimes
S_2$ space, where the Pauli spin operators and the identity that act
as a basis are, respectively, $\{\hat{I}_1,\hat{\sigma}_{x,y,z}\}$
and $\{\hat{I}_2,\hat{\rho}_{x,y, z}\}$. In this description, ${
P}^0(\vec{q}, \Omega) $ becomes:
 \begin{eqnarray}
 P^0 (\vec{q}, \Omega)& = &\sum_{\sigma}\left[P^0_{\sigma,\sigma}(\vec{q}, \Omega)\frac{\hat{I }_1+
 \sigma\hat{\sigma}_z}{2} \otimes \frac{\hat{I}_2 +
 \sigma\hat{\rho_z}}{2} \right.\nonumber \\
 & & \left.+ P^0_{\sigma,-\sigma}(\vec{q}, \Omega)\frac{\hat{I}_1 +
 \sigma\hat{\sigma_z}}{2} \otimes \frac{\hat{I}_2
 -\sigma\hat{\rho_z}}{2}\right]\;. \label{eq:matrix-prop}
 \end{eqnarray}

 The coefficients
$P^0_{\sigma,\sigma'}(\vec{q}, \Omega)$ are calculated from
Eq.~(\ref{eq:prop0}) for the corresponding pairs of spins in the
perturbative approach of the weak scattering approximation, by
considering $1/\tau_0 \gg$ much larger than all the other
frequencies involved:
\begin{eqnarray}
P^0_{\sigma\sigma} &=& 2\pi N_0\tau_0[1 + i\Omega \tau_0 -D_\sigma
q^2\tau_0]\;,\\
P^0_{\sigma,-\sigma} &=&2\pi N_0 \tau_0 [1+ i(\Omega +2\sigma \gamma
B/\hbar)\tau_0 -D_\sigma q^2\tau_0]\;. \label{eq:props}
\end{eqnarray}
 $D_\sigma = D_0(1+\sigma\zeta)$ are the spin dependent diffusion
coefficients, which in the case of the unpolarized gas become equal
to the diffusion coefficient in 2D, $D_0 = v_F^2\tau_0/2$ , where
$v_F$ is the Fermi energy.

The introduction of SOI changes substantially this picture because
the spin direction is no longer parallel to the applied magnetic
field, but rather undergoes a process of randomization induced by
the coupling to the orbital motion. A single-electron Green function
that incorporates both interactions can be obtained by using the
exact eigenfunctions and the corresponding eigenvalues of $H_{SOI}$.
Because of the spin mixing, however, this algorithm is laden with
serious mathematical difficulties, especially where the impurity
average is concerned.

In a different approach, adopted here, we exploit the linear
coupling of the electron momentum to the spin in the Rashba term.
This coupling allows the introduction of a spin-dependent vector
potential $m\tilde{\lambda}(\vec{\sigma}\times \hat{y})/\hbar$ that
is being used as a generator of a non-abelian unitary transformation
in the spin space to produce the Green's function of the
spin-polarized electrons in the presence of the spin-orbit
interaction. Such an approximation has been discussed before in
connection with a magnetic vector potential \cite{altshuler}, with
the Aharonov-Casher effect \cite{mathur}, and with SOI in parallel
quantum dots \cite{falko}. Consequently, we define
\begin{equation}
{\cal G}^{R,A}_{\vec{p},\omega}(\vec{r},\vec{r'}) =
e^{-i{\lambda}(\vec{\sigma}\times \hat{y})\cdot \Delta \vec{r}}
G^{R,A}(p,\omega)\; \label{eq:soi-green}
\end{equation}
as the single particle Green function in the presence of SOI, with
$\lambda= \tilde{\lambda}m/\hbar $ a reduced Rashba coupling
constant. On account of SOI, the new Green function ${\cal G}$ is
not diagonal. Accordingly, the non-averaged propagator ${\cal
P}^0(\vec{q},\Omega)$, defined by Eq.~(\ref{eq:prop0}), becomes a
sixteen term sum in the vector space $S_1\otimes S_2$, that reflects
the skew-scattering of the electron spins.

Performing the statistical average over the impurity configuration
requires the propagator ${\mathcal P}(\vec{q},\Omega)$ to satisfy:
\begin{equation}
{\mathcal P}(\vec{q},\Omega) = u^2+u^2{\mathcal
P}^0(q,\Omega){\mathcal P}(\vec{q},\Omega)\;. \label{eq:dyson}
\end{equation}
 An elegant solution
to Eq.~(\ref{eq:dyson}) is obtained in a basis of the $S_1 \otimes
S_2$ vector space in which the the kernel ${\mathcal
P}^0(\vec{q},\Omega)$ assumes a diagonal form. The vectors
$\Psi_i(\vec{r})$ of this basis are found by solving
\begin{equation}
\int d\vec{r'}{\mathcal P}^0(\vec{r},\vec{r'})\Psi_i(\vec{r'}) =
\Lambda_i \Psi_i(\vec{r}) \;, \label{eq:eigenprop}
\end{equation}
for the corresponding eigenvalues $\Lambda_i$.
 In this representation, $
{\cal P}^0(\vec{r},\vec{r'}) = \sum_{i}\Lambda_i (\vec{q},\Omega)
\Psi_i (\vec{r})\Psi_i^*(\vec{r'})\;
$
leading to an immediate solution to Eq.~(\ref{eq:dyson}):
\begin{equation}
{\cal P} (\vec{r},\vec{r'})=
\sum_{i}\frac{u^2}{1-u^2\Lambda_i(\vec{q},\Omega)}\Psi_i
(\vec{r})\Psi_i^*(\vec{r'})\;.
\end{equation}
The logarithmic corrections to the conductivity are then given by
the poles of ${\mathcal P}(\vec{q},\Omega)$, estimated in the limit
of small $\vec{q}$: $\lim_{\vec{q}\rightarrow 0}(1- u^2\Lambda_{i})$
\cite{altshuler}. The problem is centered, therefore, on solving the
eigenfunction-eigenvalue equation of the non-averaged propagator.

 Eq.~(\ref{eq:eigenprop}) is linearized by expanding $P^0_{\sigma\sigma'}$ and $\Psi_i(\vec{r'})$
up to second order around $\vec{r}$ \cite{altshuler}. The final
result involves the Fourier transforms of the propagators in the
absence of SOI, $P^0_{\sigma\sigma'}$, defined in
Eq.~(\ref{eq:props}), and their derivatives with respect to $q$.
Simultaneously, the eigenfunction $\Psi_i(\vec{r'})$ is replaced by
$\Psi_i(\vec{r'})=\Psi_i(\vec{r})+ \nabla \Psi_i(\vec{r})\Delta
\vec{r} + \frac{1}{2}\nabla^2 \Psi_i(\vec{r})(\Delta \vec{r})^2$.
$\Psi_i(\vec{r})$ is a spinor in $S_1 \otimes S_2$  whose general
form is a linear combination of eigenvectors of $S_{z i}$, in
$S_1\otimes S_2$, $|\xi>_i$ for spin up and $|\eta>_i$ for spin down
$(i=1,2)$ : $ \Psi (\vec{r}) =
e^{\vec{q}\cdot\vec{r}}(a|\xi>_1|\xi>_2 + b
|\xi>_1|\eta>_2+c|\eta>_1|\xi>_2 + d|\eta>_1|\eta>_2) \; $, where
$a, b, c, d$ are numerical coefficients. With this test solution,
the characteristic equation for the eigenvalues of the propagator is
obtained to be
\begin{equation}
\left| \begin{array}{cccc} Z_{1\uparrow} & q_z & q_z &
-\lambda \\
q_z & Z_{0\uparrow} & -\lambda & q_z \\
q_z &-\lambda &Z_{0\downarrow} & q_z \\
-\lambda & q_z & q_z & Z_{1\downarrow} \end{array} \right|=0 \;,
\end{equation}
where the diagonal coefficients that incorporate the eigenvalues
$\Lambda$ are $Z_{1\sigma} = \left[ 1+ i\Omega \tau_0 - D_\sigma
\tau_0 \left(q^2-4\lambda q_x - 6\lambda^2\right)
 - \Lambda \right]/2\lambda D_\sigma \tau_0$ and
$Z_{0\sigma} = \\\left[ 1+ i\left(\Omega +2\sigma\gamma B/\hbar
\right) \tau_0 - D_\sigma \tau_0\left(q^2 - 2\lambda^2\right) -
\Lambda \right]/2\lambda D_\sigma \tau_0\; $. $q_x$ and $q_z$ are
the corresponding components of $\vec{q}$.

The straightforward evaluation of the determinant leads to a quartic
equation in $\Lambda$ (imbedded in $Z_{i\sigma}$):
\begin{eqnarray}
& &\left(Z_{0\uparrow}Z_{0\downarrow}-\lambda^2\right)
\left(Z_{1\uparrow}Z_{1\downarrow}-\lambda^2\right)=\nonumber\\
 & &
q_z^2\left (Z_{1\uparrow}+Z_{1\downarrow}+2\lambda\right)
\left(Z_{0\uparrow}+Z_{0\downarrow}+2\lambda\right)\;.
 \label{eq:major}
\end{eqnarray}
Even though, in principle, this equation can be solved for any
values of $q$, for the problem at hand suffices to obtain its
solutions for $q=0$.  It is also instructive to investigate the
solutions of this equation in the absence of the magnetic field. In
the relevant limit $\vec{q}\rightarrow 0$, we obtain $Z_{1\pm} =
\pm\lambda$ and $Z_{0\pm} = \pm\lambda$, which generate the
following values for the logarithmic corrections to the
conductivity:
\begin{equation}
\lim_{q\rightarrow 0}\frac{1-u^2\Lambda_i}{\tau_0} =
\left\{\begin{array}{l}
-i\Omega + 8\lambda^2 D_0 \\
-i\Omega + 4 \lambda^2 D_0\\
-i\Omega + 4\lambda^2 D_0\\
-i\Omega
\end{array} \right.\;. \label{eq:pikus}
\end{equation}
By using the definitions of the diffusion coefficient and of the
reduced Rashba interaction, we identify $4D_0\lambda^2 =
2\tilde{\lambda}^2 k_F^2\tau_0$ ($k_F$ is the Fermi momentum) as the
Dyakonov spin relaxation rate, $1/\tau_s(0)$ \cite{dyakonov}. This
is associated with the SOI induced rate-of-change  of the in-plane
spin components : $dS_{i}/dt = -S_i/\tau_{ii}, (i=x,z)$, where
$\tau_{ii} = \tau_s(0)$. The rate of change of the perpendicular
spin component, $dS_y/dt$ is described by $\tau_{yy} =\tau(0)/2$.
The imaginary frequency is replaced by the dephasing rate,
$\tau_\phi^{-1}(0)$. With these substitutions, we readily regain the
traditional form of the conductivity poles expressed in terms of the
spin-relaxation times as derived in Ref. \cite{pikus}. One can also
show that the eigenfunction $\Psi(r)$ associated with the eigenvalue
$\tau_\phi^{-1}(0)$ corresponds to the pairing of two electrons of
opposite spins and momenta, i.e. the singlet Cooperon, and generates
the antilocalization correction \cite{pikus}.

Inspired by this analysis, we rewrite Eq.~(\ref{eq:major}) in terms
of a new set of unknowns,  $u_{i,\sigma} = Z_{i,\sigma} + \lambda $,
such that
\begin{equation}
\left[
\frac{u_{1\uparrow}u_{1\downarrow}}{u_{1\uparrow}+u_{1\downarrow}}-\lambda\right]\left[
\frac{u_{0\uparrow}u_{0\downarrow}}{u_{0\uparrow}+u_{0\downarrow}}-\lambda\right]
=q_z^2\;
\end{equation}
In this form, the limit $q\rightarrow 0$ can be taken and the
resulting two uncoupled quadratic equations can be solved
independently. With $\mu_{1\sigma} =2\pi N_0 \tau_0[ 1 + i\Omega
\tau_0 - 4\lambda^2 D_\sigma \tau_0 ]$ and $\mu_{0\sigma}=2\pi
N_0[1+i(\Omega + 2\gamma \sigma B/\hbar)\tau_0 -2\lambda^2D_0
\tau_0]$, the solutions are:
\begin{eqnarray}
\Lambda_{i,\pm} & = &
\frac{1}{2}\left[\mu_{i\uparrow}+\mu_{i\downarrow} -
2\lambda^2(D_{\uparrow}+D_{\downarrow})\tau_0 \pm \right.\\
\nonumber & & \left.
\sqrt{\left[\left(\mu_{i\uparrow}-\mu_{i\downarrow}\right)-
2\lambda^2\tau_0^2(D_{\uparrow}-D_{\downarrow})\right]^2 + 16
\lambda^4 \tau_0^4 D_\uparrow D_\downarrow}\right]
\end{eqnarray}
 A quick inspection shows that $1-u^2\Lambda_{1,\pm}\left|_{\vec{q}=0}\right./\tau_0$ and
  $1-u^2\Lambda_{0,-}\left|_{\vec{q}=0}\right./\tau_0$
 correspond to the first three lines of Eq.~(\ref{eq:pikus}). Their
 dependence on the magnetic field is producing a negligible
 effect on the logarithmic corrections to the conductivity compared
 with the unpolarized case.

The opposite is true, however, about the effect of the magnetic
field on the solution that describes the weak antilocalization,
\begin{eqnarray}
&&\lim_{q\rightarrow 0}\frac{1-u^2\Lambda_{0,+}}{\tau_0} = -i\Omega
+2D_0\lambda^2 -
\nonumber\\
& &2D_0\lambda^2 \sqrt{1+ 2\zeta\left(\frac{i\gamma B}{\hbar
D_0\tau_0\lambda^2}\right) - \left(\frac{\gamma B}{\hbar
D_0\lambda^2\tau_0}\right)^2}\;.
\end{eqnarray}
In analogy with the unpolarized case, described by the last line of
Eq.~(\ref{eq:pikus}), a dephasing time $\tau_\phi (B)$ is defined as
a measure of the antilocalization correction in the presence of a
magnetic field:
\begin{eqnarray}
& & \frac{1}{\tau_\phi(B)} = \frac{1}{\tau_\phi (0)} +
2D_0\lambda^{2} \nonumber\\
& &\Re e \left[1 - \sqrt{1+2\zeta \left(\frac{i\gamma B}{\hbar
D_0\tau_0\lambda^2}\right)+\left(\frac{i\gamma
B}{\hbar D_0\tau_0\lambda^2}\right)^2}\right]\;.\nonumber\\
& &  \label{eq:diff}
\end{eqnarray}
[$\Re e$ designates the real part of the expression.]

\begin{figure}[h]
\begin{center}
\includegraphics[width=2.5in]{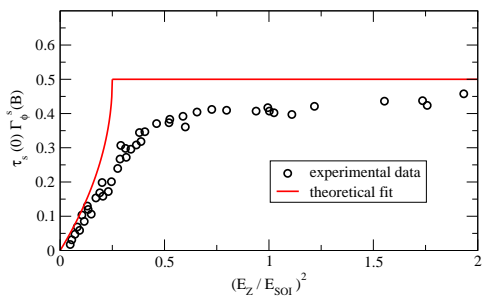}
\end{center}
\vspace*{0.05cm} \caption{ The calculated dephasing time from
Eq.~(\ref{eq:final}) is compared with the experimental data of
Meijer et al.~\cite{meijer} for $\zeta=0$.} \label{fig:1}
\end{figure}

 Following the
notations of Ref.~\onlinecite{meijer}, we introduce the variation of
the dephasing rate from the unpolarized case, $\Gamma_\phi^s(B) =
[\tau_\phi(B)]^{-1}-[\tau_\phi (0)]^{-1}$ and recognize that
$\left({\gamma B}/{\hbar D_0 \tau_0 \lambda^2}\right)^2 =\{4\gamma
B/[\hbar/ \tau_s(0)]\}^2=(2E_Z/E_{SOI})^2 $. Moreover, since on
account of the large electron density and the low effective mass the
spin polarization is very small, in the following considerations
$\zeta$ is set to $0$. Thus, by extracting the real part of
Eq.~(\ref{eq:diff}) we obtain that the change of the dephasing time
induced by the longitudinal field is simply:
\begin{eqnarray}
& & \tau_s(0)\Gamma_\phi^s (B) =\nonumber \\  & &
\frac{1}{2}\left\{1-
\sqrt{\frac{1-\left(\frac{2E_Z}{E_{SOI}}\right)^2+
\left|1-\left(\frac{2E_Z}{E_{SOI}}\right)^2\right|}{2}}\right\}\;.\label{eq:final}
\end{eqnarray}
We compare the relative dephasing rate give by Eq.~(\ref{eq:final})
with the experimental data in Fig.~(\ref{fig:1}).

The dephasing mechanism suggested by Eq.~(\ref{eq:final}) can be
understood by considering that in the presence of the longitudinal
magnetic field the down-spin component of the singlet Cooperon is
forced to undergo a spin-flip. This action is counterbalanced by the
spin-orbit coupling which will redirect the spin in some other
direction. For the singlet Cooperon to continue to exist and thus
contribute to the antilocalization, the dephasing of the original
down-spin electron trajectory has to be such that, as a result of
the spin-orbit coupling, the electron spin reverts to the down
position in respect to the direction of the field. The two
consecutive spin flips imply the $2E_Z$ energy dependence.

At weak fields, when $(2E_Z/E_{SOI})^2<<1$, from
Eq.~(\ref{eq:final}), we obtain
\begin{equation}
\tau_s(0)\Gamma_\phi^s(B) = \left(\frac{E_Z}{E_{SOI}}\right)^2 \;.
\label{eq:tau}
\end{equation}
This is equivalent with $\tau^{-1}_\phi(B) - \tau^{-1}_\phi (0) =
2\gamma^2 B^2/(k_F^2\tau_0^2)$, the result of Ref.~\cite{malshukov}.
For $(2E_Z/E_{SOI})^2 \simeq 1$, Eq.~(\ref{eq:final}) becomes:
\begin{equation}
\tau_s(0)\Gamma_\phi^s(B) =
\frac{1}{2}\frac{\left(\frac{2E_Z}{E_{SOI}}\right)^2}{1+
\sqrt{1-\left(\frac{2E_Z}{E_{SOI}}\right)^2}}\;,
\end{equation}
establishing the onset of the saturated behavior, that continues
through the high field limit, when the dephasing rate is close to
$1/2\tau_s(0)$ as observed experimentally \cite{meijer}. In this
regime, at a given intensity of the spin-orbit coupling $E_{SOI}$,
there is a maximum value of the longitudinal magnetic field
determined by $2E_{Zmax} = E_{SOI}$ for which the modifications
induced in the direction of one of the spins in the singlet Cooperon
can be compensated by the skew-scattering produced by SOI, such that
the singlet pairing is preserved. Increasing the intensity of the
magnetic field beyond this limit, leads to the disappearance of the
antilocalization correction as the singlet Cooperon configuration is
not realized any more, exactly as observed experimentally
\cite{meijer}. The dephasing of the two trajectories remains
constant since it is established solely by the spin-orbit coupling
at a value equal to half the spin-relaxation rate along the $z$
direction, $(2\tau_s(0))^{-1}$, $2\tau_s(0)$ being the amount of
time in which the original down-spin flips and then realigns itself
with the $\hat{z}$ axis. In the ideal 2D system modeled here, no
surface effects have been considered. Consequently, this calculation
does not reproduce the slow slope of the curve registered
experimentally \cite{meijer}.

We conclude, therefore, that a general theory of localization
effects in spin-polarized 2DEGs explains the saturated universal
dependence on $(2E_Z/E_{SOI})^2$ of the electron-dephasing time,
matching the data of Meijer et al.~\cite{meijer}.

\begin{acknowledgments}
I would like to thank G. E. W. Bauer for many illuminating
discussions. This research was supported in part by the National
Science Foundation under Grant No. PHY99-07949.

\end{acknowledgments}

\end{document}